\documentclass[twocolumn]{aastex63}
\usepackage{natbib}
\usepackage{graphicx}
\begin{document}
\title{Conceptual remote distance measurement with a double-slit interference}

\author[0000-0002-5400-3261]{Yuan-Chuan Zou}
\affiliation{School of Physics, Huazhong University of Science and
Technology, Wuhan, 430074, China. Email: zouyc@hust.edu.cn }
\affiliation{Perimeter Institute for Theoretical Physics, Waterloo, Ontario N2L 2Y5, Canada}

\begin{abstract}
Distance measurement is crucial to astronomy. Here we suggest a new conceptual method to measure the distance by using a local instrument. By engaging the double-slit interference and by considering the phase information of the light, the position of the intensity maximum is related to the distance of the source. Consequently, the precise measurement of the position can be used to measure the distance of the remote source.
\end{abstract}



\section{Introduction}
\label{intro}
The precise distance of an astronomical object is important to all aspects of astronomy, such as to the Hubble constant tension \citep{2019arXiv190710625V}. The cosmic distance ladder depends crucially on the direct distance measure, which mainly refers to the annual trigonometric parallaxes. With parallaxes, Gaia already measured the distances of 1.33 billion stars in our Milky Way galaxy \citep{2018AJ....156...58B}. This method mainly suffers the angular resolution of the telescopes \citep{fleisch_kregenow_2013}.

The phase information of the electromagnetic waves has been used in observations to enhance the angular resolution, such as Very Long Baseline Array (VLBA) in radio band \citep{1994IEEEP..82..658N} and Very Large Telescope Interferometer (VLTI) in optical band \citep{2000SPIE.4006....2G}. By digging into the phase information from the source, here we suggest an alternative method to directly measure the distance, which also uses the trigonometric information of the source as well as a double-slit.  We describe the concept in \S \ref{sec:con}, discuss the improvement of the concept in \S \ref{sec:imp}, and conclude in \S \ref{sec:dis}.

\section{Conceptual Configuration}
\label{sec:con}

\begin{figure}
  \includegraphics[width=0.5\textwidth]{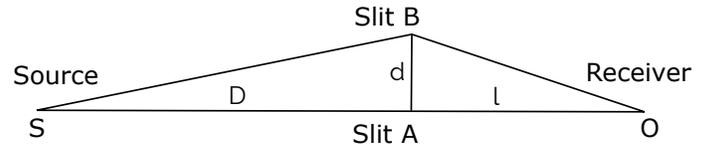}
\caption{A schematic configuration for the remote distance measurement using double-slit interference. The two slits are marked as Slit A and Slit B, which are perpendicular to the plot. The light starts from the source (point S), passes through the double slits, and interferes at point O, where the receiving instrument locates. The three points S, A and O are on the same line. The double-slit line AB is perpendicular to the line SAO. The distance from the source to the slits is $D$, from slits to the receiver is $l$, and from Slit A to Slit B is $d$. The point O is set as the first order fringe maximum of the interference of the two light beams, i.e., the difference between the two light paths is the wavelength of the light. By measuring $l$, one can infer the distance of the source $D$.}
\label{fig:sketch1}
\end{figure}

This method is based on the phase information of the electromagnetic wave from the source, while the phase information can be extracted by the double-slit interference. A schematic configuration is shown in figure \ref{fig:sketch1}. The monochromatic light from the source splits into two beams passing through Slits A and B. The two beams meet at point O, where a receiver is located to measure the intensity of the light. We can set this point the first order fringe maximum of the interference, i.e., one wavelength difference of the two beams. This point can be determined by moving the receiver back and forth, where the intensity maximum is found. Therefore, slits A and B and the receiver are the local instrumental system, while the source is at remote waiting for distance measurement.

Considering the distance of the two slits, $d$ is significantly smaller than the distance of the source $D$ and the distance between the slits to the receiver $l$, one can obtain the relation between the two beams of light as follows,
\begin{equation}
\label{eq:1}
 \frac{d^2}{2 D} + \frac{d^2}{2l} = \lambda,
\end{equation}
where $\lambda$ is the wavelength of the light emitted from the source. As $\lambda$, $d$ and $l$ are all measurable in principle, one can get the distance of the source $D$ from equation (\ref{eq:1}). {To perform the measurement, one can move the receiver back and forth to locate the position of intensity maximum, i.e., to get the value of $d$.}
The difference between this system and Young's double-slit interferometer is that the lengths of the two beams from the source to Slits A and B are not the same, which is the key point to measure the distance $D$.

The distance $D$ can be astronomical distance if $l$ can be measured very precisely. By varying $D$ and $l$ of equation (\ref{eq:1}), one can get how precisely of $l$ should be to measure a source in astronomical distance:
\begin{equation}
\label{eq:deltal}
\delta l = - \frac{l^2 \delta D}{D^2}.
\end{equation}
With 1\% precision of $D$, i.e., $\frac{\delta D}{D} = 1\%$, an astronomical distance $D = 10^{20}$m ($\sim 1$ kpc), the source emitting at the optical band (V band, $\lambda \simeq 5.5 \times10^{-7}$ m), and setting $d=0.1$ m, these derive $l \simeq 9090$ m from equation (\ref{eq:1}).
Taken all these numbers into equation (\ref{eq:deltal}), we get $\delta l  \simeq 8.1 \times 10^{-15}$ m. That is, with the precision of the position measurement of the receiver up to $\sim 10^{-14}$ m, one can measure the distance up to 1 kpc with precision in the order of 1\%.
Considering the precision could be up to $10^{-19}$ m in the LIGO experiment \citep{1992Sci...256..325A}, it is in principle achievable to measure the astronomical distances.
One of the challenges is that the system should be in the size of 10 km.

{ 
On the other hand, to locate the precise position of the fringe peak, a high signal-to-noise ratio (SNR) is required. Assuming that the peak intensity is 1, and the intensity profile obeys cosine function, the intensity at $l+\delta l$ can be represented by $\cos(2\pi \delta l/l)$  $\sim 1-(2\pi \delta l/l)^2/2$. So the required SNR is $2/(2 \pi \delta l/l)^2$, which is around $6\times 10^{34}$. Considering the photon number to be accumulated is SNR$^2$, this is another challenge that requires to accumulate an extremely large amount of photons.
}

\section{Improvement of the System}
\label{sec:imp}
\begin{figure}
  \includegraphics[width=0.5\textwidth]{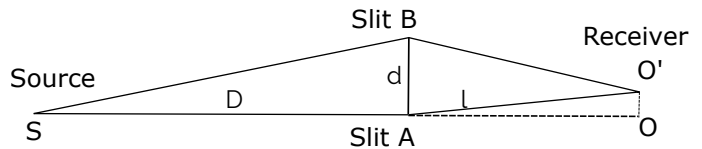}
\caption{A schematic configuration similar to figure \ref{fig:sketch1}. The only difference is that the receiver is set to point O$'$, which is close to the bisection of AB. OO$'$ is perpendicular to line SAO. $l$ is still the length of AO.}
\label{fig:sketch2}
\end{figure}

The most challenge of the configuration shown in figure \ref{fig:sketch1} is that the size of the system should be too large if one tends to measure astronomical distances. In fact, it is not necessary to set the receiver (point O in figure \ref{fig:sketch1}) on the same line of SA. One can set point O' close to the bisection of AB, as shown in figure \ref{fig:sketch2}, which is less straightforward in concept.
The distance between O and O' can be set as $\frac{d}{2}(1-\epsilon)$, where $\epsilon$ is a small number.

The relation of the two light paths reads
\begin{equation}
\label{eq:3}
 \frac{d^2}{2 D} + \sqrt{l^2+\left(\frac{d}{2}\right)^2(1+\epsilon)^2 } -  \sqrt{l^2+\left(\frac{d}{2}\right)^2(1-\epsilon)^2 } = \lambda.
\end{equation}
If we set $\epsilon=10^{-3}$, with the same other parameters, $\frac{\delta D}{D} = 1\%$, $D = 10^{20}$m, $\lambda \simeq 5.5 \times10^{-7}$ m, and $d=0.1$ m, one gets $l \simeq 4.55$ m, which is much smaller than the configuration shown in figure \ref{fig:sketch1}.
Similar to equation (\ref{eq:deltal}), by varying $D$ and $l$ of equation (\ref{eq:3}) and assuming $d \ll l$, one can get how precisely of $l$ is needed to measure a source in astronomical distance:
\begin{equation}
\label{eq:deltal2}
\delta l \simeq - \frac{l^4 \delta D}{D^2 d^2 \epsilon^2}.
\end{equation}
This requires the $\delta l \simeq 4.2 \times 10^{-12}$ m with same parameters above. { The size of the system} is also much easier to achieve. { The required SNR is around $6\times 10^{22}$ instead, which is still too high to achieve.}

There are also other aspects that could enhance the system.
To further reduce the size of the system, one may engage the resonant cavity to increase light path on the way of AO and BO.
The optical grating system might also be considered to get a higher contrast between intensity maxima and minima. {  For a number of $N$ slits/grooves, the width of the fringe peak is proportional to $1/N$. With $N \sim 10^4$, it makes the required SNR 4 orders of magnitude smaller.}

{ To avoid the movement of the receiver for finding the fringe peaks, we can use a plate placing perpendicular to the line of sight, which could be at the position OO$'$ in figure \ref{fig:sketch2}. In this case, the fringe peaks shift on the plate if the sources locate at different distance $D$. Suppose the zeroth order fringe peak is at $d_0$, and the first order fringe is at $d_1$, and so on, where $d_i$ is the distance of the ith-order fringe peak to the line SAO. We have the following relations:
\begin{equation} \label{eq:di}
 \frac{d^2}{2 D} + \sqrt{l^2+(d-d_i)^2 } -  \sqrt{l^2+d_i^2 } = i \lambda,
\end{equation}
where $i$ indicates the ith order fringe peak. The change of source distance causes the shift of $d_i$ with the following relation:
\begin{equation} \label{eq:ddi}
   \delta d_i = \frac{d^2}{2D} \frac{ \delta D}{D} \left[ \frac{d-d_i}{\sqrt{l^2+(d-d_i^2)}} + \frac{d_i}{\sqrt{l^2+d_i^2}} \right]^{-1}.
\end{equation}
In the limit $l \gg d$, it can be simplified as $\delta d_i = \frac{ld}{2D} \frac{ \delta D}{D}$.
With the same system settings as in section \ref{sec:con}, one gets $\delta d_i \simeq 4.5 \times 10^{-20}$ m, which is, however, too tiny to detect.
}

\section{Conclusion and Discussion}
\label{sec:dis}
In this work, we proposed a conceptual system to measure the distance of a source in astronomical distance. This system consists of two slits and a receiver. The light starts from the remote object, travels through the double-slit, and the receiver measures the light intensity of the interference from the double-slit. By locating the position of the first fringe maximum, one can infer the distance of the remote object.
The key feature of this system is using the phase information of the light from the source. The main difference from  Young's double-slit experiment is that the light lengths of the two paths between the source to the slits are different. This difference carries information of the distance.  However, to measure astronomical distance, it is a big challenge on how to identify the tiny shift of the fringe peak.

We also showed several aspects to enhance the system, which can be used to make the system smaller and more precise.
The advantage of the system is that, similar to the trigonometric parallax systems, it is a geometric measure, and is not dependent on any assumptions about the physical properties of the objects.
Not like the annual trigonometric parallaxes, which are based on the telescopes' (together with the Earth) orbital positions around the Sun, this system measures the distance of an object simultaneously. It has the potential to measure the distance even beyond kpc.
Besides precisely measuring the position of the receiver, precisely setting the direction of the double-slits is also challenging. The direction of AB in figure \ref{fig:sketch1} is not necessarily perpendicular to the line of sight. It could be slightly larger than $90^\circ$, which can make the system even more compact. However, one should have the exact value of that angle.
Notice this is only a conceptual design, and there are difficulties such as the required extremely SNR.
The real performable system must need many technical enhancements.
For example, telescopes may be engaged to collect more light for the interference, and if it is necessary, not only the first fringe maximum but other maxima can be used to localize its position.
{However, even with these enhancements, i.e., similar configuration as shown in figure \ref{fig:sketch2}, together with telescope collecting more photons and gratings shrinking the fringe peaks, for instance, with an object at distance $D= 10^{20}$ m, brightness of 10 mag, band width of 50 nm, total photon collecting area of 100 m$^2$, and with 10$^4$ groove gratings, one still needs $\sim 9\times 10^{30}$ s accumulation time to obtain the required SNR. It is far from achievable, not saying the bandwidth itself may smear the fringe peak. Therefore, significantly advanced design has to be engaged to overcome the difficulties in measuring the astronomical distance.}

A table instrument can be first constructed to check the feasibility and to find further difficulties might encounter. According to equations (\ref{eq:1}) and (\ref{eq:deltal}), a conceptual table experiment can be performed by setting $D=10$ m, $d=0.001$ m, which yields $l=1$ m,  $\delta l = 0.001$ m, and the required SNR is $\sim 5.1 \times 10^4$. These are all easily achievable in principle.  However, one should be cautious that with much smaller $D$, the light path difference might exceed the wavelength even at slits A and B, and the maximum of the intensity should be multiple wavelengths difference instead. {With the successful performance of the table experiment, one may be able to step to further distance and to figure out the enhancement of the instrumentation.} {A routine for the achievement of the remote distance measurement could be first a table instrument measuring the manual light source, second the sun, and then further to planets and nearby stars. A routine for the technical enhancements could be first engaging the grating system, then using telescopes to collect more photons, and further seeking more efficient way to locate the fringe peaks.}

\acknowledgements
 I thank the anonymous referee for the critical comments.
 Research at Perimeter Institute is supported by the Government of Canada through the Department of Innovation, Science and Economic Development Canada and by the Province of Ontario through the Ministry of Economic Development, Job Creation and Trade. This work was supported by the National Natural Science Foundation of China (Grant No. U1738132 and 11773010).

\bibliographystyle{aasjournal}
\bibliography{bib}

\end{document}